\documentclass[
	aps,prl,
	reprint,
	amsmath,amssymb,
	groupedaddress,
	]{revtex4-2}
\usepackage{graphicx}
\usepackage[hidelinks]{hyperref}
\usepackage{microtype}
\usepackage{lmodern}
\usepackage{sourcesanspro}
\usepackage{dakmath}

\begin{document}

\title{Extreme wave skewing and dispersion spectra of anisotropic elastic plates}
\author{Daniel A. Kiefer}
\email[]{daniel.kiefer@espci.fr}
\author{Sylvain Mezil}
\author{Claire Prada}
\affiliation{Institut Langevin, ESPCI Paris, Université PSL, CNRS, 75005 Paris, France}
\date{\today}

\begin{abstract}\noindent
Guided wave dispersion is commonly assessed by Fourier analysis of the field along a line, resulting in frequency-wavenumber dispersion curves. In anisotropic plates, a point source can generate multiple dispersion branches pertaining to the same modal surface, which arise due to the angle between the power flux and the wave vector. We show that this phenomenon is particularly pronounced near zero-group-velocity points, entailing up to six contributions along a given direction. Stationary phase points accurately describe the measurements conducted on a monocrystalline silicon plate.
\end{abstract}

\maketitle 

Guided elastodynamic waves in thin structures have applications for microelectromechanical sensors and filters~\cite{yantchevThinfilmZerogroupvelocityLamb2011,caliendoZerogroupvelocityAcousticWaveguides2017}, 
nondestructive testing and structural health monitoring~\cite{duanInvestigationGuidedWave2019,wangGroupVelocityCharacteristic2007} 
as well as material characterization~\cite{webersenGuidedUltrasonicWaves2018,bochudRealtimeAssessmentAnisotropic2018,thelenLaserexcitedElasticGuided2021,ponschabSimulationBasedCharacterizationMechanical2019}, 
Many of the materials involved in these applications are elastically anisotropic due to their underlying microstructure. This is the case for crystals and composite laminates, but also for many polycrystalline materials because their grain orientation distribution is nonuniform. It is widely acknowledged that it is crucial for the above-mentioned applications to understand the impact of anisotropy on guided wave propagation.

A complication arising in anisotropic media is the fact that power flux is not necessarily collinear with the wave vector, a well-known effect denoted as power flux skewing~\cite{auldAcousticFieldsWaves1990a,langenbergUltrasonicNondestructiveTesting2012}. 
Often studied in the context of crystals, the effect is well understood for bulk waves~\cite{everyDeterminationElasticConstants1990,everyPhononFocusingModeconversion1990} and for surface waves~\cite{maznevAnisotropicEffectsSurface2003}. For a review refer to~\cite{everyBulkSurfaceAcoustic2013}. However, the dispersive nature of guided waves in plates leads to additional complexity~\cite{auldAcousticFieldsWaves1990,royerElasticWavesSolids2022}. 
In this context, Velichko and Wilcox~\cite{velichkoModelingExcitationGuided2007}, Chapuis et al.~\cite{chapuisExcitationFocusingLamb2010} and Karmazin et al.~\cite{karmazinStudyTimeHarmonic2013} have successfully described the arbitrary point-to-point transmission of plate waves, thereby being interested in the spatial analysis regarding wave packet skewing and energy focusing.  
While these works were all concerned with the two-dimensional (2D) wave field in the plane of the plate, Glushkov et al.~\cite{glushkovGroupVelocityCylindrical2014} were interested in the field along a line away from a point source and they demonstrated that it can be explained in terms of cylindrical guided waves. Although incomplete, 1D data is very interesting in practice as it avoids the time-consuming spatial 2D scan. 

Recently, we have observed power flux skewing which covers 360$^\circ$ in the region close to zero-group-velocity (ZGV) points, i.e., where the power flux of a mode vanishes~\cite{kieferBeatingResonancePatterns2023a}. This includes transverse-group-velocity (TGV) waves, whose power flux is orthogonal to the wave vector. In this region, one mode will usually lead to multiple contributions in the wave field. Multiplicity is a well-known effect for bulk and guided waves in anisotropic media~\cite{auldAcousticFieldsWaves1990a, karmazinStudyTimeHarmonic2013,duanInvestigationGuidedWave2019,maznevAnisotropicEffectsSurface2003}, which is usually attributed to concave regions of the wavevector surface. We will see that this is different in the vicinity of ZGV points, where multiplicity is natural and does not require a concave dispersion surface. Consequently, the extreme skewing and multiplicity occurs, in principle, with arbitrarily weak anisotropy. 

In this contribution, we study the effect of power flux skewing on 1D scans, which also  explains the influence of the source on the acquired data. Using the well-established concept of ``stationary phase points'', we explain the link between the 2D dispersion surface and the measured 1D \emph{frequency-wavenumber} dispersion data. In contrast to previous work, our analysis is completely done in terms of frequency-wave vector spectra of \emph{plane guided waves}. This approach captures the mentioned multiplicity in the experimental dispersion curves. The model is validated against measurements on a monocrystalline silicon wafer, a rather weakly anisotropic  material with a universal anisotropy index of $\text{AU} = 0.24$~\cite{ranganathanUniversalElasticAnisotropy2008}. Our measurements reveal strongly pronounced skewing effects, which supports our claim that the degree of anisotropy plays a minor role.

\begin{figure}[tb]\centering\sffamily\small%
	\setlength{\fboxsep}{0pt}
	\includegraphics{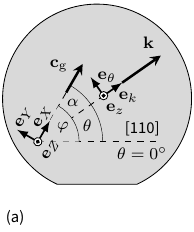}%
	\hspace{1em}%
	\includegraphics{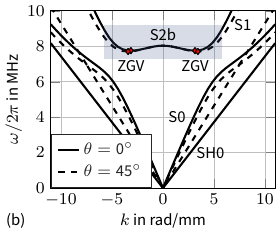}%
	\caption{Guided wave propagation in a monocrystalline silicon wafer. 
	\textbf{(a)}~Geometry in top view. Angles are defined with respect to the [110] crystal axis. $\theta$:~wave-vector orientation, $\varphi$:~observation angle, $\alpha$:~skew angle.
	\textbf{(b)}~Dispersion curves of symmetric modes for $\theta = \text{0}^\circ$ and $\theta = \text{45}^\circ$. The highlighted region of interest encompasses the S1 and S2b modes close to the ZGV points (red marks).}
	\label{fig:geometry_and_dispCurves}
\end{figure}

The theory of guided wave propagation in anisotropic plates is recalled shortly. The waves are characterized by their angular frequency~$\omega$ and wave vector~$\ten{k} = k \dirvec{k}(\theta) = k_X \dirvec{X} + k_Y \dirvec{Y}$, see Fig.~\ref{fig:geometry_and_dispCurves}a. The Cartesian system $\dirvec{X}\dirvec{Y}\dirvec{Z}$ is fixed to the plate, while $\dirvec{k}\dirvec{\theta}\dirvec{z}$ is a local system oriented with the wave vector. Taking the point of view of the wave, $\dirvec{k}$ is denoted as the \emph{axial direction} and $\dirvec{\theta}$ as the \emph{transverse direction}. Only certain combinations of $\omega$ and $\ten{k}$ can propagate. The \emph{dispersion relation}~$\omega(\ten{k})$ forms surfaces in the Cartesian $k_X$-$k_Y$-plane or, equivalently, in the cylindrical $k$-$\theta$-plane. It is usual to plot cuts across these surfaces for a chosen orientation~$\theta$ of the wave vectors~\cite{thelenLaserexcitedElasticGuided2021,bochudRealtimeAssessmentAnisotropic2018,webersenGuidedUltrasonicWaves2018,duanInvestigationGuidedWave2019}. Fig.~\ref{fig:geometry_and_dispCurves}b shows the dispersion curves along the [110] and the [010] axes of a [001]-cut monocrystalline silicon plate with 525\,{\textmu}m thickness (adjusted to 524.6\,{\textmu}m to better match the measurements to be presented). The material's stiffness is of cubic anisotropy (Voigt-notated stiffness $C_{11} = \mathup{165.6\,GPa}$, $C_{12} = \mathup{63.9\,GPa}$, $C_{44} = \mathup{79.5\,GPa}$, mass density $\rho = \mathup{2330\,kg/m^3}$). 
The angle $\theta$ is commonly referred to as \emph{propagation direction}. But for a lossless waveguide, the propagation of a wave packet is described by the group velocity vector~$\ten{c}_\mup{g}$, which might not be oriented at $\theta$. The \emph{group velocity} is given by~\cite{langenbergUltrasonicNondestructiveTesting2012,auldAcousticFieldsWaves1990}
\begin{equation} \label{eq:group_vel}
	\ten{c}_\mup{g} = \nabla_\ten{k} \omega = \frac{\partial \omega}{\partial k} \dirvec{k} + \frac{1}{k} \frac{\partial \omega}{\partial \theta} \dirvec{\theta} \,,
\end{equation}
and is proportional to the wave's power flux. When $\ten{c}_\mup{g} \cdot \ten{k} < 0$, we speak of a \emph{backward wave} and this is the case for the S2b mode. 

In isotropic media, the second term in (\ref{eq:group_vel}) vanishes and the group velocity is collinear to the wave vector, which justifies the notion of propagation direction. The situation is different in anisotropic plates. While $\ten{k}$ is oriented at angle~$\theta$, $\ten{c}_\mup{g}$ is at angle~$\varphi$. The difference, $\alpha = \varphi - \theta$, is denoted as \emph{steering} or \emph{skew angle} \cite{langenbergUltrasonicNondestructiveTesting2012,chapuisExcitationFocusingLamb2010,karmazinStudyTimeHarmonic2013}. The angle $\varphi$ dictates the observability of the corresponding wave component. To explain this, assume a source at the origin of $\dirvec{X}\dirvec{Y}\dirvec{Z}$ and a point of interest at $X \dirvec{X}$, where $\dirvec{X}$ is the \emph{observation direction}. Waves with group velocity $\ten{c}_\mup{g}$ oriented along $\dirvec{X}$ are denoted as \emph{stationary phase points}~\cite{chapuisExcitationFocusingLamb2010,glushkovGroupVelocityCylindrical2014,karmazinStudyTimeHarmonic2013} and only these waves contribute to the field at $X \dirvec{X}$ (disregarding evanescent waves). In fact, expanding the wave field with this set of propagating modes is equivalent to a stationary phase approximation of the far field~\cite{chapuisExcitationFocusingLamb2010}. 

\begin{figure}[tb]\centering\sffamily\small%
	\includegraphics{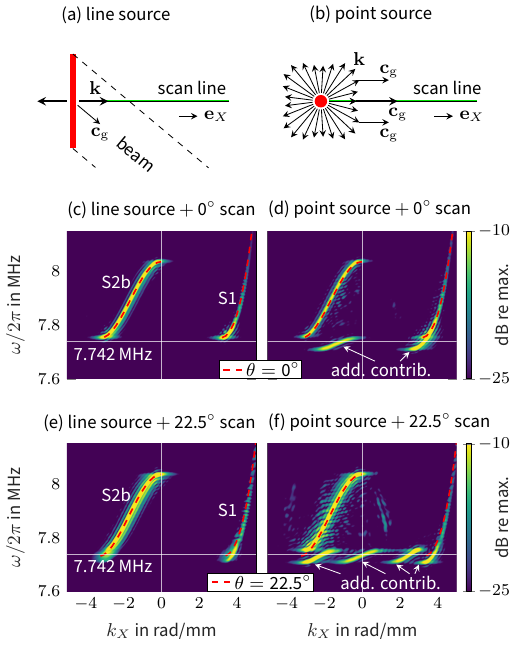}
	\caption{Line source vs. point source on an anisotropic plate. \textbf{(a)}~The line source excites (mostly) two wave vectors and the emitted beam is skewed. \textbf{(b)}~The point source excites a broad spectrum of wave vectors and some of them are observable along the scan line. \textbf{(c,d)}~Corresponding spectra of the out-of-plane surface displacements measured along the [110] axis of a silicon plate. \textbf{(e,f)}~Same as (c,d) but scanned at 22.5$^\circ$ out of the [110] axis.}
	\label{fig:line_vs_point}
\end{figure}

Our goal is to model and identify waves measured on a line away from a source. Which waves are observed depends on the source in two ways: (i) the wave-vector spectrum that it excites, and (ii) its aperture, which defines a range of observation directions for each point on the scan line. To understand the differences between the extreme cases of a point source and a line source, we sketch the excitation of a single mode at a given frequency in Fig.~\ref{fig:line_vs_point}.
The line source excites mostly two wave vectors collinear with the observation direction. 
Although the skewed power flux inclines the radiated beam, the wave is nonetheless observed on the scan line thanks to the extend (aperture) of the source. 
As expected, the measured wavenumber will be $k_\mup{X} = \ten{k} \cdot \dirvec{X} = k$ because $\dirvec{k} = \dirvec{X}$. This means that the acquired dispersion data is correctly explained by the theoretical dispersion curves of Fig.~\ref{fig:geometry_and_dispCurves}b. Because of the finite length of the source, the wave can only be measured within a section of the scanned line, as seen in Fig.~\ref{fig:line_vs_point}a. Hence, we expect a skew angle-dependent broadening of the acquired wavenumber spectrum. 

Measurements are performed with a laser-ultrasonic setup similar to Ref.~\cite{kieferBeatingResonancePatterns2023a}. It consists of a pulsed laser source (10-ns duration, 1064-nm wavelength) and an interferometer (532-nm wavelength) that measures the surface normal displacements. The source forms either a $\approx$7\,mm-long line or a small spot on the plate's surface, with 4.5\,mJ and 3\,mJ energy, respectively. Both sources excite wavenumbers up to $k\approx$~20\,rad/mm. The scan is performed on a 40\,mm-long line away from the source by displacing the interferometer with a translation stage in steps of 0.1\,mm. The temporal acquisition is 100\,{\textmu}s long. A spatio-temporal Fourier transform with a 20\%-tapered cosine window in both dimensions yields the spectra depicted in Fig.~\ref{fig:line_vs_point}c-\ref{fig:line_vs_point}f. With a line source, both the scan along $\varphi = \text{0}^\circ$ (Fig.~\ref{fig:line_vs_point}c) and the one at $\varphi = \text{22.5}^\circ$ (Fig.~\ref{fig:line_vs_point}e) show excellent agreement to the dispersion curves computed for $\theta = \text{0}^\circ$ and $\theta = \text{22.5}^\circ$, respectively. Note that $\theta = \text{0}^\circ$ is a symmetry axis and the corresponding skew angles are zero, while this is not the case for $\theta = \text{22.5}^\circ$. The mentioned skew-dependent broadening is confirmed when comparing Fig.~\ref{fig:line_vs_point}e to Fig.~\ref{fig:line_vs_point}c. 

The point source sketched in Fig.~\ref{fig:line_vs_point}b excites wave vectors in all possible directions~$\theta$. Due to anisotropy, these wave vectors~$\ten{k}$ exhibit different magnitudes and associated group velocities~$\ten{c}_\mup{g}$. Certain specific modes will have $\ten{c}_\mup{g}$ directed along the scan line and are, hence, measured. The experimentally acquired wavenumbers are the projection of these wave vectors onto the observation direction, i.e., $k_X = \ten{k} \cdot \dirvec{X} = k \cos{\alpha}$. This explains the discrepancy in Fig.~\ref{fig:line_vs_point}f with respect to the superposed dispersion curves computed for $\theta = \text{22.5}^\circ$, in particular, close to the minimum in frequency. Most remarkably, multiple $\ten{k}$ might propagate energy in the chosen observation direction, leading to the mentioned multiplicity. As seen in Fig.~\ref{fig:line_vs_point}d and \ref{fig:line_vs_point}f, this manifests in additional dispersion branches that also pertain to the S1 and S2b modes. 

\begin{figure}[tb]\sffamily\small%
  \includegraphics{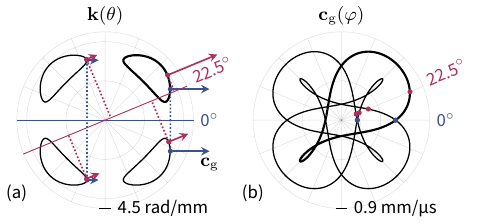}
  \caption{Cuts at 7.742\,MHz of \textbf{(a)}~the dispersion surface and \textbf{(b)} the ray surface. 
  The modes observable at 0$^\circ$ and 22.5$^\circ$ are marked therein. The wave vector projections (dotted line) onto the corresponding direction (solid line) were measured in Fig.~\ref{fig:line_vs_point}d and \ref{fig:line_vs_point}f.
  }
  \label{fig:raySurf}
\end{figure}

While the sketch in Fig.~\ref{fig:line_vs_point}b is qualitative, in Fig.~\ref{fig:raySurf} we plot the actual isofrequency contour $\ten{k}(\theta)$ and ray contour $\ten{c}_\mup{g}(\varphi)$ at 7.742\,MHz. Note that $\ten{c}_\mup{g}$ is always normal to the isofrequency contours. For both measurement directions, we marked the corresponding four stationary phase points, i.e., plane waves that propagate energy in the observation direction. They explain the additional contributions in Fig.~\ref{fig:line_vs_point}d and \ref{fig:line_vs_point}f. Note that only two of the solutions are actually observed for the 0$^\circ$-scan because two pairs of points exhibit the same $k_X$. The number of stationary phase points varies with the scan angle. Fig.~\ref{fig:raySurf}b exhibits six of them in a narrow range close to 45$^\circ$. This is due to the small loops in the ray contours that can be attributed to the concave regions in the dispersion contours. In the following, we analyze the frequency-dependence of the acquired response while restricting to the case of a point source. The results are applicable to any source distribution via Fourier analysis.

\begin{figure*}[tb]\sffamily\small%
  \includegraphics{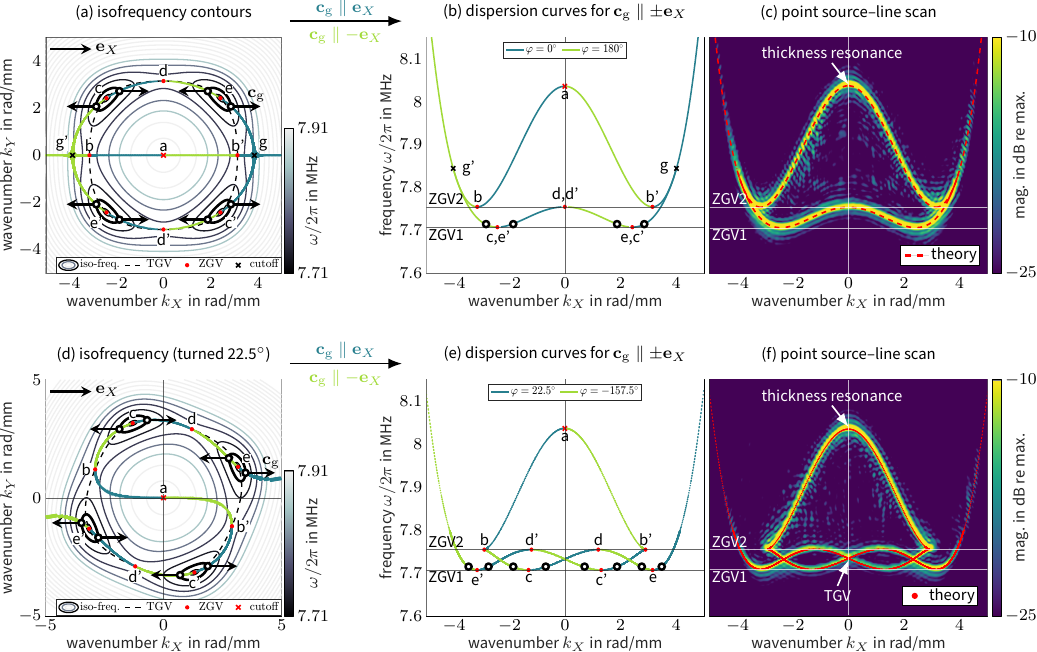}
  \caption{Waves observable with a point-source and a line-scan along $\dirvec{X}$. The top row \textbf{(a-c)} is for a scan along the [110] axis ($\varphi = \text{0}^\circ$), while the bottom row \textbf{(d-f)} is for a scan at angle $\varphi = \text{22.5}^\circ$. \textbf{(a,d)}~Dispersion surface of the S1/S2b modes as iso-frequency contours oriented such that the observation direction $\dirvec{X}$ is horizontal. The stationary phase points marked therein in dark (light) green are the waves observable to the right (left) of the source. Observable group velocity vectors are sketched for one selected frequency. \textbf{(b,e)}~Stationary phase points as frequency~$\omega/2\pi$ vs. horizontal wavenumber~$k_X$. The highlighted points from (a,d) are also marked. \textbf{(c,f)}~Measured spectral magnitude obtained by scanning across the point source vs. the theoretical stationary phase points from (b,e).}
  \label{fig:silicon0deg}
\end{figure*}

The full dispersion surface of the S1 and S2b modes is depicted in the form of iso-frequency contours in Fig.~\ref{fig:silicon0deg}a. The modes exhibit four ZGV points on $\theta = 45^\circ + n \times 90^\circ$, $n \in \mathbb{Z}$ ($\langle 100 \rangle$ axes) that correspond to minima of the dispersion surface and are denoted as ZGV1 (marked as e, e', c and c')~\cite{kieferBeatingResonancePatterns2023a}. Additionally, it exhibits four ZGV points on $\theta = n \times 90^\circ$ ($\langle 110 \rangle$ axes) that correspond to saddle points (b, b', d, d'). The latter are at a slightly higher frequency and are denoted as ZGV2. For $\theta$ outside symmetry axes, the minimum in the dispersion curve corresponds to a TGV wave, i.e., its power flux is orthogonal to the wave vector. The dashed curve in Fig.~\ref{fig:silicon0deg}a marks the loci of TGV waves~\cite{kieferBeatingResonancePatterns2023a}. 

A point source excites modes on the entire dispersion surface. The loci of the wave vectors that can be measured right (left) to the source are marked on the dispersion surface of Fig.~\ref{fig:silicon0deg}a as dark (light) green points. As $\ten{c}_\mup{g}$ is normal to the isofrequency contours, these stationary phase points correspond to the locations where the contours are vertical. The horizontal [110]-axis is a reflection symmetry axis that coincides with the observation direction and all points on this line are stationary phase points. Additional stationary phase points are found close to the TGV waves. First, the isofrequency contours close to a ZGV1 point (minimum) encircle this point away from the origin. This implies one off-axis point where $\ten{c}_\mup{g}$ is oriented along $\dirvec{X}$ and one where it is along $-\dirvec{X}$, as is indicated on the highlighted convex contours. Second, the ZGV2 points (saddle points) induce regions where the surface is concave (e.g., between b' and g). Therefore, the ZGV2 point at $0^\circ$ ($180^\circ$) leads to two symmetric stationary phase points for observation along $+\dirvec{X}$ ($-\dirvec{X}$). The ZGV points and the thickness resonance at $k = 0$ are always stationary phase points and it is here that the group velocity $\ten{c}_\mup{g}$ changes direction.

Projecting the stationary phase points onto the $\omega$-$k_X$-plane and plotting them as conventional frequency-wavenumber dispersion curves, leads to the representation in Fig.~\ref{fig:silicon0deg}b. We observe two curves in ``W''-shape. While the upper ``W'' is due to the points on the horizontal axis of the dispersion surface of Fig.~\ref{fig:silicon0deg}a, the lower one is due to the points encircling the center. We remark that only the projection of the latter onto the observation direction is measured, meaning a-priori that the wave vector magnitude $k$ remains unknown. As the lower and the upper stationary phase points of the circle in Fig.~\ref{fig:silicon0deg}a have the same $k_X$, they merge in the lower ``W'' curve of Fig.~\ref{fig:silicon0deg}b. This manifests in coalescence of the eight stationary phase points marked in Fig.~\ref{fig:silicon0deg}a into four points in Fig.~\ref{fig:silicon0deg}b. We remark that the lower ``W'' has an upper cutoff frequency marked by a cross. It is given by point g on the dispersion contours and above this frequency the contours become purely convex. In contrast to Fig.~\ref{fig:line_vs_point}, we now scanned \emph{across the source} with a distance of 40\,mm to each side. This is why we obtain the symmetric spectrum ($\varphi = 0^\circ$ and $\varphi = 180^\circ$) seen in Fig.~\ref{fig:silicon0deg}c. The theoretical predictions are superposed on the measurements and the agreement is remarkably good.

Next, to measure along a line that is not a symmetry axis, we rotate the plate by -22.5$^\circ$ while keeping the observation directions $\pm \dirvec{X}$ horizontal. The stationary phase points for the rotated material are marked on the dispersion surface of Fig.~\ref{fig:silicon0deg}d. They form an ``S''-shaped path. Fig.~\ref{fig:silicon0deg}e shows the corresponding frequency-wavenumber dispersion forming twisted curves with up to eight different wavenumbers per frequency (or up to three frequencies per wavenumber). These points are highlighted for one frequency in Fig.~\ref{fig:silicon0deg}d and \ref{fig:silicon0deg}e. Four of them pertain to the S1 mode (outside the TGV curve), while the other four are associated to the S2b mode (inside the TGV curve). Like previously, the waves observable to the right (left) correspond to the branches with positive (negative) slope in Fig.~\ref{fig:silicon0deg}e. The corresponding experimental dispersion curves are depicted in Fig.~\ref{fig:silicon0deg}d and are seen to coincide again with the theoretical predictions. 

To the best of our knowledge, no algorithm has been devised to directly compute the stationary phase points. As a work-around, we have computed the dispersion surface of the S1/S2b modes on a dense set of about 2.1 million ($k, \theta$)-points. Subsequently, we selected the solutions where $\ten{c}_g$ is in the desired direction within 2\,mrad ($\approx$0.1$^\circ$) tolerance. The computation was done with our open source software \texttt{GEWtool}~\cite{kieferGEWtool2023} and takes about 1.5\,min using eight cores of an Apple M1 Pro processor. 

In conclusion, the frequency-wavenumber spectra obtained by scanning a line away from a source on an anisotropic elastic plate is explained by the stationary phase points on the dispersion surface. We showed that ZGV points entail the unfolding of a mode into multiple well-resolved contributions in the experimentally acquired spectra. This was validated by measurements on a monocrystalline silicon wafer, a rather weakly anisotropic material. Furthermore, the extreme power flux skewing induced by the ZGV points implies that wave vectors spanning 360$^\circ$ contribute to the data acquired in a single direction away from a point source. This phenomenon opens perspectives for the inverse characterization of complex anisotropic materials with limited data, as it inherently contains directional information. Similar phenomena are expected in more complex systems, such as metamaterials~\cite{bossartExtremeSpatialDispersion2023} and polariton propagation in anisotropic media~\cite{zhangUltrafastAnisotropicDynamics2023}. 

\medskip
\begin{acknowledgments}
The authors are thankful for helpful comments by Daniel Royer. This work has received support under the program ``Investissements d’Avenir'' launched by the French Government under Reference No. ANR-10-LABX-24.
\end{acknowledgments}

\bibliography{main}

\end{document}